\documentclass[12pt,pra,aps,preprintnumbers,showpacs,superscriptaddress]{revtex4}
\usepackage{amssymb}
\usepackage{epsfig}

\newcommand{\be}{\begin{equation}}
\newcommand{\ee}{\end{equation}}
\newcommand{\bea}{\begin{eqnarray}}
\newcommand{\eea}{\end{eqnarray}}

\newcommand{\beq}{\begin{equation}}
\newcommand{\eeq}{\end{equation}}
\newcommand{\la}{\langle}
\newcommand{\ra}{\rangle}
\begin{document}
\title{Solitons in Maximally Entangled Two Qubit Phase Space}
\date{\today}
\author{O. K. \surname{Pashaev}}
      \affiliation{Department of Mathematics,
Izmir Institute of Technology
Urla-Izmir, 35430, Turkey}
\author{Z. N. \surname{Gurkan}}
\affiliation{Center for Quantum Technologies, National University of Singapore, 117542, Singapore}

\begin{abstract}
Motivated by M\"obius transformation for symmetrical points under the generalized circle in complex plane,
the system of symmetrical spin coherent states corresponding to antipodal qubit states is introduced. It implies
the maximally entangled spin coherent states basis, which in the limiting cases reduces to the Bell basis.
A specific property of our symmetric image coherent states is that they never become unentangled for any value of $\psi$ from complex plane.
By the reduced density matrix and the concurrence determinant methods, it is shown that our basis is maximally
entangled. In addition we find that the average of spin operators in these states vanish, as it must be according to another,
operational definition of completely entangled states. Universal one qubit and two qubit gates in this new basis are calculated and time evolution of these states for some spin systems is derived. We find that the average energy for XYZ model in two qubit case (Q symbol of H) shows regular finite energy localized structure with characteristic extremum points, and appears as a soliton in maximally entangled two qubit phase space.
Generalizations to three and higher qubit states are discussed.
\end{abstract}
\maketitle
\section{Introduction}
The coherent states has been introduced first by E. Schr\"odinger for the harmonic oscillator, in attempt to construct
maximally classical states evolving according to classical equations of motion \cite{Schrodinger}. Then R.J. Glauber considered
the coherent states on the Heisenberg-Weyl group \cite{Glauber}, which becomes important tool for description of coherent
laser beams in quantum optics. Generalized  coherent states to arbitrary Lie group has been invented by A. Perelomov
and some particular realizations have been discussed by many researchers \cite{Perelomov1}.
Important class of coherent states for SU(2) and SU(1,1) groups
describes spin waves in Heisenberg spin model of ferromagnetism \cite{PMS}, \cite{MP1}.
The SU(1,1) coherent states has been applied also in  superfluidity theory for description of the Bogoliubov condensate state \cite{Solomon},
in terms of  pseudospin \cite{MP2}, \cite{MP1}. Impossible here to mention all papers devoted to the subject.
But in all of these applications the direct product of coherent states is explored, which means separability of the states.

Recently in quantum information and quantum computation theory the entangled coherent states become
interesting tool to study entanglement in quantum systems. This way in addition to the Glauber coherent state $|\alpha\ra$ another
state $|-\alpha\ra$ has been considered. In terms of these states entangled states like $|\alpha\ra |\alpha\ra + |-\alpha\ra|-\alpha\ra$
has been discussed by several authors \cite{Barry}. However, these states are not orthogonal
$ \la -\alpha| \alpha \ra =e^{-2 |\alpha|^2}$, and it creates several complications of
computational and interpretational character. Thought one can construct the even and odd coherent states \cite{Dodonov}
$(|\alpha\ra \pm |-\alpha\ra)/\sqrt{2}$, which are orthonormal and thus determine the qubit basis in two dimensional Hilbert space.

In the present paper, motivated by M\"obius transformation and its action on symmetrical points of the generalized circle in complex plane, we introduce the set
of spin $1/2$ coherent states which are orthogonal and maximally entangled. The paper is organized as follows. In Sec. 2 we pedagogically introduce relation between M\"obius transformation and qubit. Special attention we paid for so called symmetrical points in unit circle, which appear in the method of images from hydrodynamics, and related symmetric qubit quantum states. In Section 3 we construct an orthonormal basis from symmetric antipodal qubit states and elementary gates as M\"obius transformations. Section 4 devoted to symmetric two qubit coherent states. In Section 5 we show that the set of states introduced in previous section is maximally entangled. We follow two different methods, the reduced density matrix method and the determinant method. Then we show that the average of spin operators in our states is vanish. This property also confirm that our states are maximally entangled, according to another, operational definition of entangled states as maximally non-classical states.  As an application of our results, in Section 6 we calculate average energy on our coherent states (Q symbol of H) for two and three qubit cases in XYZ model. This energy surface shows regular character with extremum  points and appears as a soliton in maximally entangled qubit space.
Time evolution of concurence and fidelity in coherent states are derived in Section 7.

\section{Linear Fractional Transformation and Symmetrical Quantum States}

\subsection{M\"obius transformation and qubit}
There is well known relation between the group of linear fractional transformations, or the M\"obius transformations,
\begin{equation}
w = S(\psi) = \frac{a \psi + b}{c \psi + d}\label{MT}
\end{equation}
$a d - b c \neq 0$
and group of two-by-two complex matrices \cite{Ahlfors}.
 Any matrix from the group acting on states
 \begin{equation}
 \left(\begin{array}{c}w_1\\ w_2 \end{array} \right) = \left(\begin{array}{cc}a & b\\ c & d \end{array} \right) \left( \begin{array}{c}\psi_1\\ \psi_2 \end{array}\right)
 \label{lintr}\end{equation}
  in terms of homogeneous coordinates
$\psi = \psi_1/\psi_2$, $w = w_1/w_2$, implies fractional transformation  (\ref{MT}). If we consider two quantum states
from two dimensional Hilbert space $| \psi \ra =\left(
                                                  \begin{array}{cc}
                                                    \psi_1 & \psi_2 \\
                                                  \end{array}
                                                \right)^T
 $ and
$| w \ra =\left(
                                                  \begin{array}{cc}
                                                    w_1 & w_2 \\
                                                  \end{array}
                                                \right)^T
 $  related by linear transformation $|w\ra = U |\psi \ra$, then it implies
fractional transformation (\ref{MT}) in extended complex plane $C$.

In quantum computations we have a qubit as a unit of information
\be
|\psi \ra = \left( \begin{array}{c}\psi_1\\ \psi_2 \end{array}\right), \,\,\,\,|\psi_1|^2 + |\psi_2|^2 = 1 \label{z}
\ee
then, in terms of homogeneous coordinate $\psi = \psi_2/\psi_1$ we have
\be
|\psi\ra = \left( \begin{array}{c}\psi_1\\ \psi_2 \end{array}\right) = \psi_1 \left( \begin{array}{c}1\\ \psi \end{array}\right).
\ee
We fix $\psi_1$ by normalization condition $\la \psi|\psi \ra = 1$, so that up to the global phase we have the
qubit state
\be
|\psi \ra = \frac{1}{\sqrt{1 + |\psi|^2}} \left( \begin{array}{c}1\\ \psi \end{array}\right).\label{cs}
\label{CS1}\ee
This state coincides with the spin $1/2$ generalized coherent state \cite{Perelomov}. From another side,
%solving normalization condition (\ref{z}) one has
%one qubit
\be |\theta, \varphi \ra=
\cos{\frac{\theta}{2}}|0\ra + \sin{\frac{\theta}{2}} e^{i \varphi}
|1\ra = \left(
          \begin{array}{l}
            \cos{\frac{\theta}{2}} \\
            \sin{\frac{\theta}{2}}\, e^{i \varphi} \\
          \end{array}
        \right)
\ee determined by point $(\theta, \varphi)$ on the Bloch sphere, and parameterized by the homogeneous variable
\be
\psi = \frac{\psi_2}{\psi_1} = \tan \frac{\theta}{2} e ^{i\phi}
\ee
determines the stereographic projection of point $(\sin \theta \cos \phi, \sin \theta \sin \phi, \cos \theta)$
on the unit sphere to the complex plane $\psi$.
Therefore the Bloch sphere considered as a Riemann
sphere for the extended complex plane $\psi$  by the stereographic projection,
determines the $SU(2)$ or the spin coherent state \be |\psi \ra =
\frac{|0\ra + \psi |1\ra}{\sqrt{1+ |\psi|^2}}. \ee The
computational basis states $|0\ra= | \uparrow \ra= \left(
                                                     \begin{array}{cc}
                                                       1 & 0
                                                     \end{array}
                                                   \right)^T
$ and $|1\ra= | \downarrow \ra= \left(
                                                     \begin{array}{cc}
                                                       0 & 1
                                                                                                          \end{array}
                                                   \right)^T$
in this coherent state representation are just points in extended complex plane  $(\Re \psi, \Im \psi)\cup \{ \infty
\}$, as $\psi=0$ and $\psi=\infty$ respectively. These points are symmetrical points under the unit circle at the origin.
\subsection{Symmetric points}
In complex analysis these two points are
symmetrical under the unit circle $ \bar{\psi} \psi = |\psi|^2=1. $
In general, two points $\psi$ and $\psi^*$ are called symmetrical with respect to the circle $C$ through
$\psi_1$, $\psi_2$, $\psi_3$ if and only if $(\psi^*, \psi_1, \psi_2, \psi_3) = \overline{(\psi, \psi_1, \psi_2, \psi_3)}$
 where the cross ratio of four points is \be (\psi, \psi_1, \psi_2, \psi_3)=\frac{(\psi-\psi_2)\,(\psi_1-\psi_3)}{(\psi-\psi_3)\,(\psi_1-\psi_2)}\label{crossratioi}.\ee
The circle here is considered in the generalized form, that includes also a line, regarded as a circle with an infinite radius.
On the Riemann sphere all generalized circles are coming from intersection of the sphere with a plane, so that if the plane passes the
north pole, the corresponding projection would be a line.
For the unit circle at the origin, we can choose $\psi_1 = -1, \psi_2 = i, \psi_3= 1 $ so that the symmetrical point of $\psi$ is $ \psi^* = 1/\bar\psi $.
It means that points $\psi$ and $\psi^*$ have the same argument and are situated on the same half line from the origin, so that if one of the point is out of the circle, the second one is inside the circle, and vice versa. Hence points $\psi = 0$ and $\psi^* = \infty$ are symmetrical points with respect to the circle.

The cross product (\ref{crossratioi}) is invariant under the M\"obius transformation, so that
if a M\"obius transformation
carries a generalized circle $C_1$ into a circle $C_2$, then it transforms any pair of symmetrical points with respect to $C_1$ into a pair
of symmetrical points with respect to $C_2$. According to this, if one considers the M\"obius transformation, mapping the unit circle to the imaginary axis (see Sect. 3.4 and fractional transformation corresponding to the Hadamard gate),
\be \psi_H = \frac{1-\psi}{1+\psi}\ee
where the last one represents the generalized circle,
then symmetric point to complex number $\psi$ is just reflection in imaginary axis: $- \bar\psi: \psi \rightarrow - \bar\psi $. For the M\"obius transformation
\be \psi_H = i\frac{1-\psi}{1+\psi}\ee
mapping the unit circle to the real axis, the symmetric point to $\psi$ is $\bar\psi: \psi \rightarrow  \bar\psi.$ The composition of symmetric points in real axis: $\psi \rightarrow \bar\psi$, in imaginary axis: $\bar\psi \rightarrow -\psi$
and then in the unit circle: $-\psi \rightarrow -\frac{1}{\bar\psi}$, produce the inverse-symmetric point
$ \psi \rightarrow - \frac{1}{\bar\psi}$.
The above symmetric points have simple meaning on the Riemann sphere:

1. $\psi$ and $\psi^* = \bar\psi$ are projections of symmetric points $M(x,y,z)$ and $M^*(x,-y,z)$

2. $\psi$ and $\psi^* = -\bar\psi$ are projections of symmetric points $M(x,y,z)$ and $M^*(-x,y,z)$

3. $\psi$ and $\psi^* = \frac{1}{\bar\psi}$ are projections of symmetric points $M(x,y,z)$ and $M^*(x,y,-z)$

4. $\psi$ and $\psi^* = -\frac{1}{\bar\psi}$ are projections of symmetric points $M(x,y,z)$ and $M^*(-x,-y,-z)$

The last one corresponds to the antipodal points on the Riemann sphere. One notice that in the above reflections *- defines anti-automorphisms considered in \cite{Fujii}.
\subsection{Symmetric qubits}
Symmetric points are important in the hydrodynamics theory and related with the so called method of images
\cite{Milne}. For point vortex in the plane bounded by the cylindrical  domain \cite{Milne} or the annular domain (canonical region for two cylinders problem)\cite{Pashaevyilmaz}, the symmetrical points represent images of the vortex. Now we like to introduce the coherent states corresponding to symmetric points,
representing symmetrical pair of qubits with remarkable properties. Then in some sense we can speak about method of images in quantum theory. For given qubit
\be |\theta, \varphi \ra=
\cos{\frac{\theta}{2}}|0\ra + \sin{\frac{\theta}{2}} e^{i \varphi}
|1\ra = \left(
          \begin{array}{l}
            \cos{\frac{\theta}{2}} \\
            \sin{\frac{\theta}{2}}\, e^{i \varphi} \\
          \end{array}
        \right)
\label{symm1}\ee
in case 1. we have the symmetric one
\be |\theta, -\varphi \ra=
\cos{\frac{\theta}{2}}|0\ra + \sin{\frac{\theta}{2}} e^{-i \varphi}
|1\ra = \left(
          \begin{array}{l}
            \cos{\frac{\theta}{2}} \\
            \sin{\frac{\theta}{2}}\, e^{-i \varphi} \\
          \end{array}
        \right)\label{symm2}
\ee
and in the second case 2. we have
\be |\theta, \pi-\varphi \ra=
\cos{\frac{\theta}{2}}|0\ra - \sin{\frac{\theta}{2}} e^{-i \varphi}
|1\ra = \left(
          \begin{array}{l}
            \cos{\frac{\theta}{2}} \\
           - \sin{\frac{\theta}{2}}\, e^{-i \varphi} \\
          \end{array}
        \right).\label{symm3}
\ee

Since the unit circle in the $\psi$ plane: $|\psi|^2 = 1$, represents equator on the Bloch sphere, then any point on upper heme-sphere projects to
the external part of the unit circle. While the lower heme-sphere is projected to internal part of the circle. It is easy to see that
if point $M (x,y,z)$ is projected to $\psi$, then reflected in equator point $M^*(x,y,-z)$ is projected to the symmetrical point
$\psi^*$. According to these two points we have "symmetric" qubit state in the case 3.

\be |\pi - \theta, \varphi \ra=
\sin{\frac{\theta}{2}}|0\ra + \cos{\frac{\theta}{2}} e^{i \varphi}
|1\ra = \left(
          \begin{array}{l}
            \sin{\frac{\theta}{2}} \\
            \cos{\frac{\theta}{2}}\, e^{i \varphi} \\
          \end{array}
        \right)
\label{symm4}\ee
These pairs of qubit states define the symmetric qubit coherent states. The corresponding points $M$ and $M^{*}$ on Bloch sphere are mirror images of each other in coordinate planes $xz$, $yz$, $xy$ respectively. This is why we can call symmetrical qubit states \ref{symm2} , \ref{symm3}, \ref{symm4} as mirror image qubits.
For every complex number $\psi$ as projection of point $(\theta, \phi)$, we have the
coherent state (\ref{CS1}). Then every symmetric point determines the symmetric coherent state.
For symmetric point $\psi^*=\bar{\psi}$,
 \be |\bar\psi \ra = \frac{|0\ra + \bar\psi |1\ra}{\sqrt{1+|\psi|^2}}\label{23}.\ee
For the point $\psi^*= - \bar{\psi}$,
 \be |-\bar\psi \ra = \frac{|0\ra - \bar\psi |1\ra}{\sqrt{1+|\psi|^2}}\label{24}\ee
For point $ \psi^* = \frac{1}{\bar{\psi}}.$
 the symmetric coherent states of qubit
is
\be
|\psi^* \ra = |\frac{1}{\bar\psi} \ra= \frac{\bar\psi|0\ra +  |1\ra}{\sqrt{1+|\psi|^2}}
 \label{scs}\ee
In the limiting case of symmetric points $\psi = 0$ and $\psi^* = \infty$ for the first two cases we have computational basis.
In the third case the reversed basis $
|\psi = 0\ra = | 1 \ra , \,\,\, |\psi^* = \infty\ra = | 0 \ra $
Now, if one has dealing with one qubit gate represented by the linear transformation  (\ref{lintr}), then it transforms the unit circle at origin
to a generalized circle in such a way that symmetrical points in the first circle transform to symmetrical points with respect to the new one.
It will define the transformation rule for symmetric qubit states. In next sections we find these M\"obius transformations related to basic
quantum gates.

\section{Antipodal Orthogonal Symmetric coherent qubit states}

\subsection{Generalized coherent state computational basis}

According to our definition of symmetric coherent states, expansion of an arbitrary qubit state in computational basis $  |\phi \ra = c_1 |0\ra +
c_2 |1\ra \label{combasisexp}$
can be considered as an expansion to specific symmetrical coherent states. Then we have natural generalization of this expansion
to arbitrary symmetrical states (\ref{23}), (\ref{24}), (\ref{scs})
\be   |\phi \ra = d_1 |\psi \ra + d_2 |\psi^* \ra \ee
considering states  $|\psi \ra $ and $|\psi^* \ra$ as a
basis. However this basis is not orthonormal due to for (\ref{scs})
\be \la \psi^* | \psi \ra = \frac{2 |\psi|}{1+ |\psi|^2} \leq 1. \ee
\be d_1= \la \psi | \phi \ra = \frac{c_1 + \bar{\psi} c_2}{\sqrt{1+|\psi|^2}},\,\,\,
d_2= \la \psi^* | \phi \ra = \frac{|\psi_1|c_1 + \frac{|\psi|}{\psi} c_2}{\sqrt{1+|\psi|^2}}  \ee
and we have
\be |\phi \ra = \frac{c_1 + \bar{\psi} c_2}{\sqrt{1+|\psi|^2}}  |\psi\ra + \frac{|\psi_1|c_1 + \frac{|\psi|}{\psi} c_2}{\sqrt{1+|\psi|^2}} |\psi^* \ra \ee
It becomes orthonormal only in specific case of computational basis when $\psi \rightarrow 0$ or $\psi \rightarrow \infty$.
In the special case when one of the points $\psi$ belongs to unit circle $|\psi| = 1$, the symmetric points coincide $\psi = \psi^*$
so that $\la \psi^* | \psi \ra = \la \psi | \psi^* \ra=1$ and we have just one state.
\subsubsection{Antipodal qubit and Inverse-symmetric basis}
The above introduced states $|\psi \ra$ and $|\psi^*\ra$ in (\ref{23}), (\ref{24}), (\ref{scs}) are not orthogonal. To have the orthogonal states for given state $|\psi \ra$
we consider the inverse-symmetric state $|-\psi^*\ra$ in the case 4. This state is represented by point $-\psi^* = -1/\bar\psi$ which is rotation of the symmetric point $\psi^*$ on angle $\pi$, and which belongs to the line through points $\psi$ and $\psi^*$. We call this point as the inverse symmetric point or inverse mirror image
and corresponding coherent state as the inverse-symmetric coherent state (inverse mirror image state). On the Bloch sphere for point $M (x,y,z)$ representing qubit
state $|\theta, \varphi \ra$,
it is given by antipodal point $-M^*(-x,-y,-z)$ corresponding to state
\be |\pi - \theta, \varphi + \pi\ra=
\sin{\frac{\theta}{2}}|0\ra - \cos{\frac{\theta}{2}} e^{i \varphi}
|1\ra = \left(
          \begin{array}{c}
            \sin{\frac{\theta}{2}} \\
            -\cos{\frac{\theta}{2}}\, e^{i \varphi} \\
          \end{array}
        \right)
\ee
which we call the antipodal qubit state. For the state (\ref{cs}) we have explicitly
\be
|-\psi^* \ra = \frac{|0\ra - \psi^* |1\ra}{\sqrt{1+|\psi^*|^2}} = \frac{|\psi||0\ra - \frac{|\psi|}{\bar\psi} |1\ra}{\sqrt{1+|\psi|^2}} \label{iscs}\ee
Up to phase this state can be written in the form
\be
|-\psi^* \ra = \frac{-\bar\psi|0\ra +|1\ra}{\sqrt{1+|\psi|^2}}  \label{iscs1}\ee

 In contrast to symmetric state (\ref{scs}), the inverse-symmetric state (\ref{iscs}) is orthogonal to $|\psi\ra$:
\be\la-\psi^* | \psi\ra = 0\ee
Then, states $|\psi\ra$ and $|-\psi^* \ra$ form the orthonormal basis so that for any state
\be   |\phi \ra = e_1 |\psi \ra + e_2 |-\psi^* \ra \ee we have
\be e_1 = \la \psi|\phi\ra = \frac{c_1 + c_2 \bar\psi}{\sqrt{1 + |\psi|^2}},\,\,\,
e_2 = \la -\psi^*|\phi\ra = \frac{- \psi c_1  + c_2}{\sqrt{1 + |\psi|^2}}
\ee
\subsubsection{Antipodal orhogonal coherent state for arbitrary representation j}
The antipodal states can be derived also for spin $j$ representation of $su(2)$ Lie algebra:
\be [J_3, J_+] = J_+, \,\,\,[J_3, J_-] = -J_-, \,\,\, [J_+, J_-]= 2J_3\ee
where $J_\pm = \frac{1}{2} (J_1 \pm iJ_2 )$, so that
\be J_+ |j, m\ra = \sqrt{(j-m)(j+m+1)}|j, m+1 \ra,\ee
\be J_- |j, m\ra = \sqrt{(j-m+1)(j+m)}|j, m-1 \ra \ee
\be J_3 |j, m\ra = m |j, m\ra \ee
where $-j \le m \le j$. The coherent state $|\psi\ra$, $\psi \in C$,  is defined by
\be |\psi \ra = \frac{1}{(1+ |\psi|^2)^j} \sum_{k=0}^{2j}\left( \frac{(2j)!}{k! (2j-k)!}\right)^{1/2} \psi^k |j, -j + k\ra\ee
For the scalar product of two coherent states after simple calculation  we have
\be \la\phi |\psi\ra = \frac{(1+ \bar\phi \,\psi)^{2j}}{(1+ |\phi|^2)^j (1+ |\psi|^2)^j}.\ee
Then orthogonality condition implies $ 1 + \bar\phi \,\psi = 0 $
or the inverse-symmetric point in the unit circle $\phi = - \frac{1}{\bar\psi}$. Representation of these coherent states in terms of unit vector ${\bf n}$
\be \la {\bf n_1} | {\bf n_2}\ra = e^{i\theta({\bf n_1},{\bf n_2})}\left(\frac{1 + {\bf n_1}{\bf n_2}}{2}\right)^j\ee
shows that the above points are antipodal points on the sphere ${\bf n_1}{\bf n_2} = -1$.
\subsection{Unitary M\"obius transformation}

From arbitrary fractional linear transformations (\ref{MT}) determined by $SL(2,C)$ matrix for quantum computations  important is a class of unitary transformations
given by  matrix
\be U= \left(
                           \begin{array}{cc}
                             a & b \\
                             - \bar b &  \bar a \\
                           \end{array}
                         \right)
 \label{unitary} \ee
  where $|a|^2 + |b|^2 = 1$.
Acting on a qubit  state $|\psi \ra = \psi_1 |0\ra +
\psi_2 |1\ra$, it implies the M\"obius transformation
\be \phi_U = \frac{a \psi + b}{- \bar b \psi + \bar a}\ee
for $\psi = \psi_1/\psi_2$ and the linear transformation
\be |\psi_U\ra     = | \frac{a \psi + b}{- \bar b \psi + \bar a}\ra\ = U |\psi\ra \ee
 acting, up to the phase,
on the coherent state
 \be
|\psi \ra = \frac{1}{\sqrt{1 + |\psi|^2}} \left(
\begin{array}{c}\psi
\\ 1 \end{array}\right)\label{notcs}. \ee This state differs from the
state (\ref{cs}) by flipping transformation.

The NOT gate $\sigma_1 = X \equiv NOT$ acts on a qubit state $|\psi \ra = \psi_1 |0\ra +
\psi_2 |1\ra$ as flipping
\be |\psi_{NOT}\ra =  \sigma_1 |\psi\ra = \psi_2 |0\ra +
\psi_1 |1\ra     \ee
and imply the M\"obius transformation $ \psi \rightarrow \psi_{NOT} = \frac{1}{\psi}$
connecting (up to phase)  two coherent states (\ref{cs}) and (\ref{notcs}):
\be
|\psi_{NOT} \ra  = \frac{1}{\sqrt{1 + |\psi_{NOT}|^2}} \left( \begin{array}{c}\psi_{NOT} \\ 1 \end{array}\right)=
\frac{1}{\sqrt{1 + |\psi|^2}} \left( \begin{array}{c} 1 \\ \psi \end{array}\right).
\ee
Due to this we denote the flipped state (\ref{notcs}) as $|\psi_{NOT}\ra$. Next we consider the Hadamard gate and the phase gate, as the universal gates.

1. The Hadamard gate
\be H= \frac{1}{\sqrt{2}}\left(
                           \begin{array}{cr}
                             1 & 1 \\
                             1 & -1 \\
                           \end{array}
                         \right)
  \ee
acts on coherent state (\ref{cs}) as
\be |\psi_H \ra = H |\psi \ra = \frac{1}{\sqrt{1+|\psi|^2}} \left(
                                                        \begin{array}{c}
                                                          1+ \psi \\
                                                          1-\psi \\
                                                        \end{array}
                                                      \right)=  \frac{1}{\sqrt{1+|\psi_H|^2}} \left(
                                                        \begin{array}{c}
                                                          1 \\
                                                          \psi_H \\
                                                        \end{array}
                                                      \right),\ee
and implies the M\"obius transformation
 \be  \psi_H = \frac{1-\psi}{1+\psi}\ee
  on state (\ref{scs})
 \be  |\psi^*_H \ra = H |\psi^* \ra = \frac{1}{\sqrt{1+|\psi^*|^2}} \left(
                                                        \begin{array}{c}
                                                          1+ \psi^* \\
                                                          1-\psi^* \\
                                                        \end{array}
                                                      \right)=  \frac{1}{\sqrt{1+|\psi^*_H|^2}} \left(
                                                        \begin{array}{c}
                                                          1 \\
                                                          \psi^*_H \\
                                                        \end{array}
                                                      \right)\ee
                                                      or
                                                      \be \psi^*_H= \frac{1-\psi^*}{1+\psi^*} = \frac{\bar \psi - 1}{\bar \psi + 1}.\ee

 Under this fractional transformation, the unit circle $|\psi|^2 = 1$ transforms to the imaginary axis in $\psi$ plane:
$\psi_H = i \Im \psi_H$, and images of symmetric points $\psi_H$ and $\psi^*_H$ are located symmetrically around this axis.
The corresponding transformed symmetric states $|\psi_H\ra$ and $|\psi^*_H\ra$ are located on the Bloch sphere, equidistantly
from the vertical plane through $\Re \psi = 0$.
For the inverse symmetric point we obtain
\be
(-\psi^*)_H = \frac{1+\psi^*}{1-\psi^*} = \frac{\bar\psi + 1}{\bar\psi - 1} = - \frac{1}{\bar \psi_H}
\ee
This formula means that points $\psi_H$ and $(-\psi^*)_H$ are inverse-symmetric points but now with respect to the unit circle at
the origin.
So the Hadamard gate transforms symmetric points with respect to unit circle to symmetric points with respect to imaginary axis,
 and the inverse-symmetric points  to
the inverse symmetric points but with respect to the unit circle.

2. For the phase shift gate
\be R_z(\theta)= \left(
                   \begin{array}{cc}
                     1 & 0 \\
                     0 & e^{i \theta} \\
                   \end{array}
                 \right)
  \ee
  we have
\be |\psi_R \ra = R_z(\theta) |\psi \ra = \frac{1}{\sqrt{1+|\psi|^2}} \left(
                                                        \begin{array}{c}
                                                          1 \\
                                                          e^{i \theta}\psi \\
                                                        \end{array}
                                                      \right)=  \frac{1}{\sqrt{1+|\psi_R|^2}} \left(
                                                        \begin{array}{c}
                                                          1 \\
                                                          \psi_R \\
                                                        \end{array}
                                                      \right)\ee
which implies simple rotation around the origin on angle $\theta: \psi_R = e^{i\theta} \psi $. The same rotation acts on symmetric and the inverse symmetric points.

\subsubsection{Coherent Hadamard Basis}
The Hadamard basis $ H|0\ra = \frac{1}{\sqrt{2}} (|0\ra+ |1\ra)\equiv |+\ra $ and $ H|1\ra = \frac{1}{\sqrt{2}} (|0\ra- |1\ra)\equiv |-\ra $
   by transformation $U|0\ra = |\psi \ra$ and $ U|1\ra = |-\psi^* \ra $
   transforms to
   \bea U|+\ra &=& \frac{1}{\sqrt{2}} (|\psi \ra + |-\psi^*\ra)\equiv |\psi_+\ra\\
   U|-\ra &=& \frac{1}{\sqrt{2}} (|\psi \ra - |-\psi^*\ra)\equiv |\psi_\ra       \eea
   This coherent state Hadamard basis is generated from the coherent states by unitary transformation
   \bea |\psi_+ \ra &=& (UHU^{-1})|\psi\ra \\
   |\psi_- \ra &=& (UHU^{-1})|-\psi^*\ra
   \eea

\section{Two Qubit Case}

\subsection{Coherent state orthonormal basis}
Here we consider the two qubit coherent state

\be |\psi_1\ra |\psi_2\ra = \frac{1}{\sqrt{1+|\psi_1|^2} \sqrt{1+|\psi_2|^2}} \left(
                                                                                \begin{array}{cccc}
                                                                                  1 & \psi_2 & \psi_1 & \psi_1 \psi_2 \\
                                                                                \end{array}
                                                                              \right)^T
\ee
By proper choice of $\psi_1$ and $\psi_2$ we can construct two qubit orthonormal coherent states basis. We find this basis in the next form
\bea |\psi \ra | \psi \ra &=& \frac{1}{1+|\psi|^2} \left(
                                             \begin{array}{cccc}
                                               1 &
                                               \psi &
                                               \psi &
                                               \psi^2
                                             \end{array}
                                           \right)^T\\
                                         | \psi \ra |-\psi^* \ra&=& \frac{1}{1+|\psi|^2} \left(
                                             \begin{array}{cccc}
                                               -\bar \psi &
                                               1 &
                                               -|\psi|^2 &
                                              \psi
                                             \end{array}
                                           \right)^T\\
  |-\psi^*\ra | \psi \ra&=& \frac{1}{1+|\psi|^2} \left(
                                                \begin{array}{cccc}
                                                  -\bar\psi &
                                                  -|\psi|^2 &
                                                  1 &
                                                  \psi
                                                \end{array}
                                              \right)^T\\
    |-\psi^*\ra | -\psi^* \ra &=& \frac{1}{1+|\psi|^2} \left(
                                                \begin{array}{cccc}
                                                  \bar\psi^2 &
                                                  -{\bar{\psi}}&
                                                  -{\bar{\psi}} &
                                                  1
                                                \end{array}
                                              \right)^T
    \eea
They form the orthonormal coherent state basis. And can be generated from the computational basis by
operator $U=\frac{1}{\sqrt{1+|\psi|^2}}\left(
                                         \begin{array}{cc}
                                           1 & -\bar{\psi} \\
                                           \psi & 1 \\
                                         \end{array}
                                       \right)
$
\bea |\psi \ra |\psi \ra &=& (U \otimes U) |00 \ra= \hat{U}_{12}|00 \ra \\
 |\psi \ra |-\psi^* \ra &=& (U \otimes U) |01 \ra= \hat{U}_{12}|01 \ra \\
 |-\psi^* \ra |\psi \ra &=& (U \otimes U) |10 \ra = \hat{U}_{12}|10 \ra\\
 |-\psi^* \ra |-\psi^* \ra &=& (U \otimes U) |11 \ra= \hat{U}_{12}|11 \ra \eea
where $ \hat{U}_{12}= U \otimes U   $.

\subsubsection{Generation of Maximally Entangled States from Coherent State Basis}
But due to separability these states are not entangled as well as the computational basis. However we can generate maximally entangled Bell states from the computational basis by using a combination of Hadamard gate and a $CNOT$ gate. We first apply the Hadamard gate to the left qubit then apply the $CNOT$ gate $C: CNOT(H \otimes I)$
\bea  C|00\ra &=& |\phi_B^+ \ra=\frac{1}{\sqrt{2}}(|00\ra + |11\ra),\,\,\,\,\,\,\,\,  C|01\ra = |\psi_B^+ \ra= \frac{1}{\sqrt{2}}(|01\ra +|10\ra) \\
 C|10\ra &=& |\phi_B^- \ra=\frac{1}{\sqrt{2}}(|00\ra - |11\ra),\,\,\,\,\,\,\,  C|11\ra = |\psi_B^- \ra=\frac{1}{\sqrt{2}}(|01\ra -|10\ra) \eea
This allows us to introduce the next set of coherent states
\bea |P_+\ra &=& (\hat{U}_{12} C \hat{U}_{12}^{-1})|\psi \psi\ra = \hat{U}_{12} C (\hat{U}_{12}^{-1}\hat{U}_{12})|00\ra    =\hat{U}_{12}|\phi_B^+ \ra\label{p1}\\
|P_-\ra &=& (\hat{U}_{12} C \hat{U}_{12}^{-1})|\psi -\psi^*\ra = \hat{U}_{12} C (\hat{U}_{12}^{-1}\hat{U}_{12})|01\ra =\hat{U}_{12}|\phi_B^- \ra  \label{p2}\\
|G_+\ra &=&(\hat{U}_{12} C \hat{U}_{12}^{-1})|-\psi^* \psi\ra = \hat{U}_{12} C (\hat{U}_{12}^{-1}\hat{U}_{12})|10\ra =\hat{U}_{12}|\psi_B^+ \ra  \label{p3}\\
 |G_-\ra &=&(\hat{U}_{12}  C\hat{U}_{12}^{-1})|-\psi^* -\psi^*\ra = \hat{U}_{12} C (\hat{U}_{12}^{-1}\hat{U}_{12})|11\ra= \hat{U}_{12}|\psi_B^- \ra
\label{p4}\eea
The set $|P_{\pm}\ra$, $|G_{\pm}\ra$ is an orthonormal set of two qubit coherent states. And any two qubit state can be expanded in this set as a basis. The concurrence formula for two qubit state expanded in Bell basis
\be |\phi\ra= s^+ |\phi_+\ra+ s^- |\phi_-\ra + h^+ |\psi_+\ra+ h^-
|\psi_-\ra \ee  is given by \be C= |s^{+^2} - s^{-^2}-h^{+^2}+h^{-^2} |. \ee
Then we can show that for the state $|\psi\ra$ expanded according to our basis
\be
   |\phi \ra = b^+ |P_+\ra + b^- |P_-\ra + c^+ |G_+\ra + c^- |G_-\ra\ee
   the concurrence formula becomes
\be C= |b^{+^2}-b^{-^2}-c^{+^2}+c^{-^2}| . \ee

\section{Maximally Entangled Orthogonal Two Qubit Coherent States}
Here we show that the set of states (\ref{p1})- (\ref{p4})
\bea  |P_{\pm}\ra &=& \frac{1}{\sqrt{2}}(|\psi\ra |\psi \ra \pm | - \psi^*\ra |-\psi^*\ra) \label{P}\\
 |G_{\pm}\ra &=& \frac{1}{\sqrt{2}}(|\psi\ra |-\psi^* \ra \pm | - \psi^*\ra |\psi\ra)\label{G}\eea
 is maximally entangled set of orthogonal two qubit states.
 First we follow the reduced density matrix approach. The density matrix for the pure states $ |P_{\pm}\ra $ in
coherent state basis is
$ \rho_P^{\pm}= |P_{\pm}\ra \la P_{\pm}|.$
%\bea \rho_P^{\pm}&=& |P_{\pm}\ra \la P_{\pm}|\\
%  &=& \frac{1}{2}(|\psi\ra |\psi\ra \la \psi| \la \psi| \pm |\psi\ra |\psi\ra \la -\psi^*| \la -\psi^*|\\
%  &\pm& |-\psi^*\ra |-\psi^*\ra \la \psi| \la \psi|+ |-\psi^*\ra |-\psi^*\ra  \la -\psi^*| \la -\psi^*|)
%   \eea
   The reduced density matrix can be written as
$ \rho_B=tr_B \, (\rho_P^{\pm})=\frac{\mathbb{I}}{2} $
%\bea \rho_B&=&tr_B \, (\rho_P^{\pm})=   \la \psi| \rho_P^{\pm} |\psi\ra + \la -\psi^* | \rho_P^{\pm} |-\psi^*\ra \\
%&=& \frac{1}{2} (|\psi\ra \la \psi |+|-\psi^* \ra \la -\psi^*|  )\\
%&=&  \frac{1}{2}\left(
%      \begin{array}{cc}
%        1 & 0 \\
%        0 & 1 \\
%      \end{array}
%    \right)
%  \eea
  so that $tr (\rho_B)=1 $ and $tr (\rho_B)^2=\frac{1}{2} $,  hence the reduced
density operator $\rho_B$ represents a mixed state. Since the
concurrence in this state is $ C= \sqrt{2(1- tr \rho_B^2)}=1 ,$ $ |P_{\pm}\ra$ are maximally entangled states.
Similar way it can be shown that $|G_{\pm}\ra$ are also maximally entangled states.
%density matrix for the pure states $  |G_{\pm}\ra $ in coherent state
%basis $\rho_G^{\pm}= |G_{\pm}\ra \la G_{\pm}|$
%\bea \rho_G^{\pm}&=& |G_{\pm}\ra \la G_{\pm}|\\
%   &=& \frac{1}{2}(|\psi\ra |-\psi^*\ra \la \psi| \la -\psi^*| \pm |\psi\ra |-\psi^*\ra \la -\psi^*| \la \psi|\\
%  &\pm& |-\psi^*\ra |\psi\ra \la \psi| \la -\psi^*|+ |-\psi^*\ra |\psi\ra  \la -\psi^*| \la \psi|)
%   \eea
%and the reduced density matrix
%\be \rho_B=tr_B \, (\rho_G^{\pm})=\frac{\mathbb{I}}{2} \ee
%\bea \rho_B&=&tr_B \, (\rho_G^{\pm})=  \la \psi| \rho_G^{\pm} |\psi\ra + \la -\psi^* | \rho_G^{\pm} |-\psi^*\ra\\
%&=& \frac{1}{2} ( |\psi\ra \la \psi |+ |-\psi^* \ra \la -\psi^*| )\\
%&=& \frac{1}{2}\left(
%      \begin{array}{cc}
%        1 & 0 \\
%        0 & 1 \\
%      \end{array}
%    \right)
%  \eea
%gives
%$tr (\rho_B)=1 $ and $tr (\rho_B)^2=\frac{1}{2} $ so it represents a mixed state. Since the concurrence is
%\be C= \sqrt{2(1- tr \rho_B^2)}=1 \ee $ |G_{\pm}\ra$ are maximally
%entangled states.
Explicitly for these states we have

  \be |P_+\ra= \frac{1}{\sqrt{2}(1+|\psi|^2)} \left(
                                    \begin{array}{c}
                                      1+\bar{\psi}^2 \\
                                      \psi-\bar{\psi} \\
                                       \psi-\bar{\psi} \\
                                     1+\psi^2 \\
                                    \end{array}
                                  \right)\,\,\,\,\,\,\, |P_-\ra= \frac{1}{\sqrt{2}(1+|\psi|^2)}  \left(
                                    \begin{array}{c}
                                     1-\bar{\psi}^2 \\
                                      \psi+\bar{\psi} \\
                                       \psi+\bar{\psi} \\
                                     -1+\psi^2 \\
                                    \end{array}
                                  \right)   \label{143} \ee

\be |G_+\ra=\frac{1}{\sqrt{2}(1+|\psi|^2)} \left(
                                                  \begin{array}{c}
                                                    -2\bar{\psi} \\
                                                    1-|\psi|^2\\
                                                     1-|\psi|^2\\
                                                     2 \psi\\
                                                  \end{array}
                                                \right)\,\,\,\,\,\,\,
|G_-\ra= \frac{1}{\sqrt{2}(1+|\psi|^2)}  \left(
                                                  \begin{array}{c}
                                                    0 \\
                                                   1+|\psi|^2 \\
                                                  - 1-|\psi|^2 \\
                                                    0 \\
                                                  \end{array}
                                                \right)
 \label{144}   \ee
which is convenient to calculate the concurrence for pure states in the determinant form $C_{12}=\left|
                                                                                                                 \begin{array}{cc}
                                                                                                                   t_{00} & t_{01} \\
                                                                                                                   t_{10} & t_{11} \\
                                                                                                                 \end{array}
                                                                                                               \right|
$, where $t_{ij}, i,j=0,1$ are coefficients of expansion for states $|\psi\ra $ in computational basis. Applying this definition to states (\ref{143}), (\ref{144}) we find that concurrence $C_{12}=1$, as we expected.

Finally if we add two spins
\be \hat{\vec{S}}_{\pm}= \hat{\vec{S}}_1 \otimes I \pm I \otimes \hat{\vec{S}}_2 \\
   .\ee
   and calculate the average of these spins in our states (\ref{P}), (\ref{G}) then we find that they vanish.
 \bea  \la P_{\pm}| \hat{S}^z_{\pm}|P_{\pm}\ra&=&0,\,\,\,\, \la P_{\pm}| \hat{S}^+_{\pm}|P_{\pm}\ra=0 \\
  \la G_{\pm}| \hat{S}^z_{\pm}|G_{\pm}\ra&=&0 ,\,\,\,\, \la G_{\pm}| \hat{S}^+_{\pm}|G_{\pm}\ra=0 \eea
  This  property has been used as an operational definition of completely entangled states in \cite{SchKlyc}, which are considered as a maximally nonclassical states.

\section{Operators and Their Q Symbols}

Using coherent states one may represent operators acting on the Hilbert space in terms of certain class of functions, which determines the operators completely and they are called operator symbols \cite{Perelomov}. For the Glauber coherent states the operator symbol $ A(\bar{\alpha}
, \beta)= \la \alpha | A | \beta \ra $ is analytical function of complex variables $\bar{\alpha}$ and $\beta$ and it is determined completely by its diagonal values, \be  A(\bar{\alpha}
, \alpha)= \la \alpha | A | \alpha \ra ,
\ee
which are called $Q$ symbol of operator $A$.

Operator symbols can be considered as functions on the phase space of a classical dynamical system. In this case coherent states may provide the naturel means for quantization and its classical correspondence. In the following we consider the Hamiltonian operator $H$ and its average $\la \psi| H | \psi\ra$ in our spin coherent states as $Q$ symbol of $H$: $Q_H(\psi)$. As it is well known if an operator is bounded, then it always has a $Q$ symbol \cite{Perelomov}, which is a value of an entire function $\mathcal{H}(\psi, \psi)=\la \psi| H | \psi\ra $. For spin model with finite number of qubits, the Hamiltonian is bounded operator, this is why its symbol always exists and is representable as a finite function. This function, the average energy in the coherent state, appears as finite energy configuration in phase space of the system and can be considered as a soliton in the phase space. Below we study the $XYZ$ spin model for two and three qubit states, and calculate $Q$ symbol of $H$, as finite average energy in the coherent state.

Here we like to stress that our two qubit coherent states are maximally entangled states and are determined by one complex $\psi$ or two real parameters. These parameters can be fixed by concrete physical requirements on minimal energy, or some constraints on fidelity, etc. And this reduction will not change entanglement of the system.
This why our $Q$ symbol of Hamiltonian appears as a finite energy soliton in maximally entangled two (three or higher) qubit phase space.

\begin{itemize}
  \item
First we consider $XXX$ model,
\be H= -J(S_1^+ S_2^- +S_1^- S_2^+ + 2   S_1^z
S_2^z) \ee where $ S_x=
\frac{S^+ + S^-}{2}, \,\,\,\, S_y= \frac{S^+ - S^-}{2i}  $ and \bea S^+ |0\ra &=& 0,\,
S^+ |1\ra= \hbar |0\ra \\  S^- |0\ra&=& \hbar |1\ra ,\,\, S^- |1\ra
=0 \\ S_z |0\ra &=& \frac{\hbar}{2}|0\ra ,\,\, S_z |1\ra=
-\frac{\hbar}{2}|1\ra .\eea

 Then we find the $Q$ symbol of this Hamiltonian in $|P_{+}\ra$ state is a constant
\be \la P_+|H|P_+\ra=-\frac{J \hbar^2}{2}.\ee

\item
For $XXZ$ model  \be H= -J(S_1^+ S_2^- +S_1^- S_2^+) + 2\Delta   S_1^z
S_2^z \ee
where $J\Delta=J_z$
we have
\be \la P_+|H|P_+\ra= \frac{-2 \hbar^2}{(1+|\psi|^2)^2} [-J (\psi - \bar{\psi})^2 + J_z(\frac{(1- |\psi|^2)^2}{2}+ \psi^2 + \bar{\psi}^2)]\ee
For $ \psi= x+ i y $
\be \la P_+|H|P_+\ra=-\hbar^2\frac{8 J y^2 + J_z[1+ 2x^2-6y^2+(x^2+y^2)^2]}{(1+x^2+y^2)^2} \ee
In Fig.1 we show the average energy surface as a function of $x,y$ with two minima extremum points.
\begin{figure}[htbp]
\centerline{\epsfig{figure=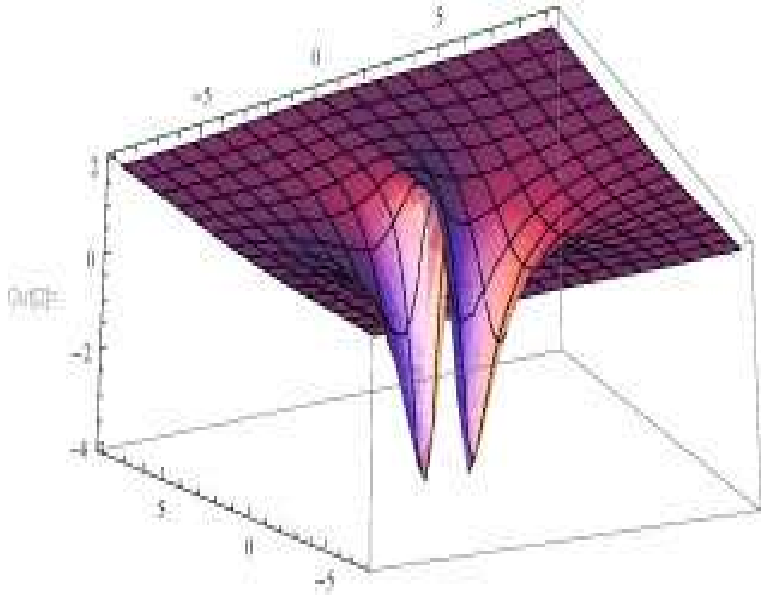
,height=8cm,width=10cm}
} \caption{$XXZ$ average energy in maximally entangled $|P_+\ra$ state for $\hbar=1, J=1, J_z=-2$}
\end{figure}
\end{itemize}
\subsection{Two Qubit case XYZ model}
Here we calculate average energy for $XYZ$ model
\be H= \frac{1}{2} [J_x \sigma_1^x \sigma_2^x + J_y \sigma_1^y \sigma_2^y + J_z \sigma_1^z \sigma_2^z ] \ee
in two qubit spin coherent states (\ref{P}, \ref{G}). In $| P_+\ra$ state we find
\be \la P_+|H|P_+\ra= \frac{- 2 J_+ (\psi - \bar\psi)^2 + J_- [(1 + \psi^2)^2 + (1 + \bar\psi^2)^2] + J_z [(1 - |\psi|^2 )^2 + 2 (\psi^2 + \bar\psi^2)]}{2 (1+|\psi|^2)^2}.\ee
In Fig.2 we show energy surface as function of $x= \Re{\psi}, y= \Im{\psi}$ with characteristic local maxima points.
\begin{figure}[htbp]
\centerline{\epsfig{figure=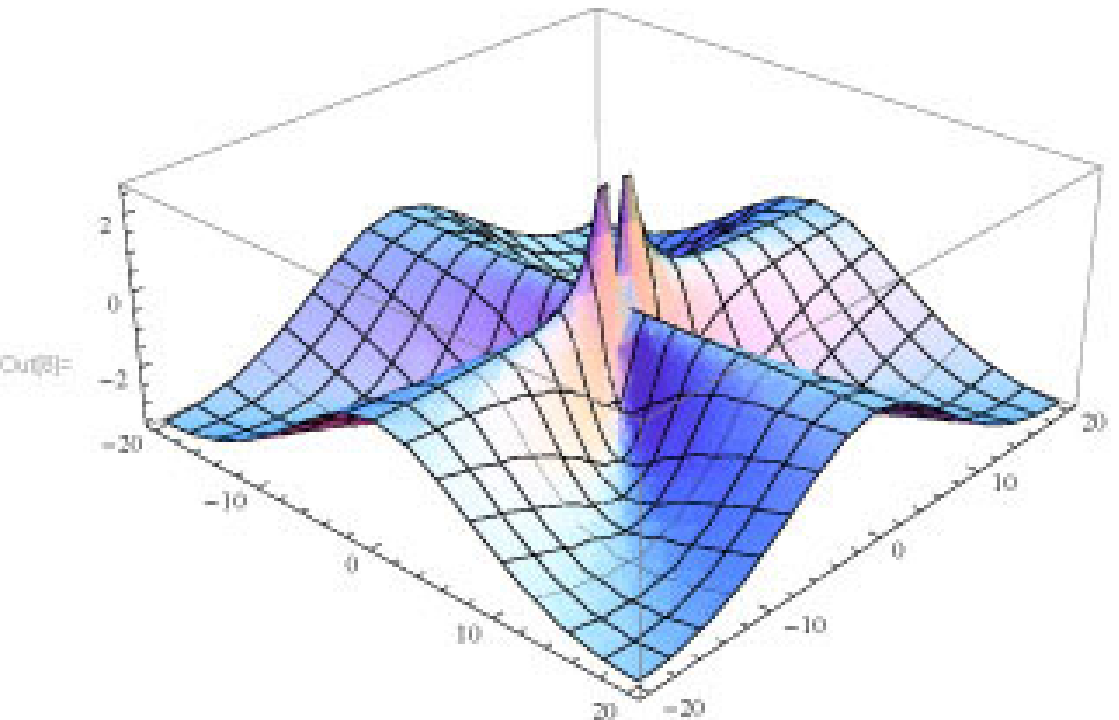
,height=8cm,width=10cm}
} \caption{$XYZ$ average energy in maximally entangled $|P_+\ra$ state for $ J_+=1, J_-= 1.5, J_z=-4$}
\end{figure}

For the state $| P_-\ra$ we have
\be \la P_-|H|P_-\ra= \frac{ 2 J_+ (\psi + \bar\psi)^2 - J_- [(1 - \psi^2)^2 + (1 - \bar\psi^2)^2] + J_z [(1 - \psi^2)(1-\bar\psi^2) - (\psi + \bar\psi)^2]}{2 (1+|\psi|^2)^2}\ee
It is shown in Fig.3.

\begin{figure}[htbp]
\centerline{\epsfig{figure=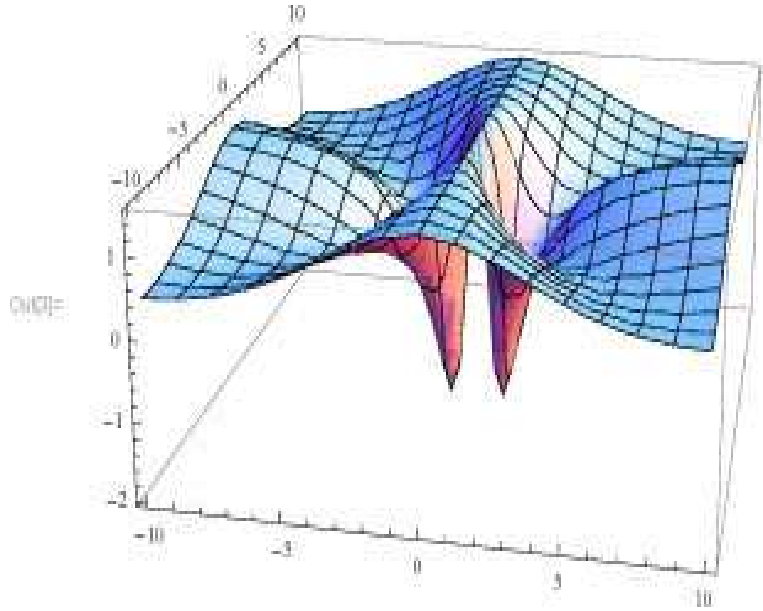
,height=8cm,width=10cm}
} \caption{$XYZ$ average energy in maximally entangled $|P_-\ra$ state for $ J_+=1, J_-= -0.5, J_z=2$}
\end{figure}

For the state $| G_+\ra$ it is
\be \la G_+|H|G_+\ra= \frac{ 2 J_+ (1- |\psi|^2)^2 - 4 J_- [\psi^2 + \bar\psi^2] + J_z [4 |\psi|^2 - (1 - |\psi|^2)^2]}{2 (1+|\psi|^2)^2}\ee
In Fig.4 and Fig. 5 we show the average energy surface for different values of parameters and specific extremum structure.
\begin{figure}[htbp]
\centerline{\epsfig{figure=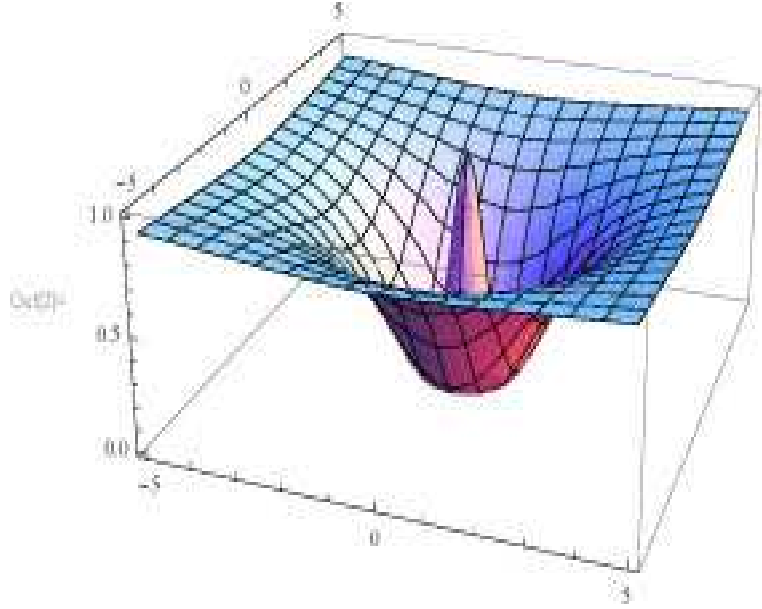
,height=8cm,width=10cm}
} \caption{$XYZ$ average energy in maximally entangled $|G_+\ra$ state for $ J_+=1, J_-= 0, J_z=0$}
\end{figure}

\begin{figure}[htbp]
\centerline{\epsfig{figure=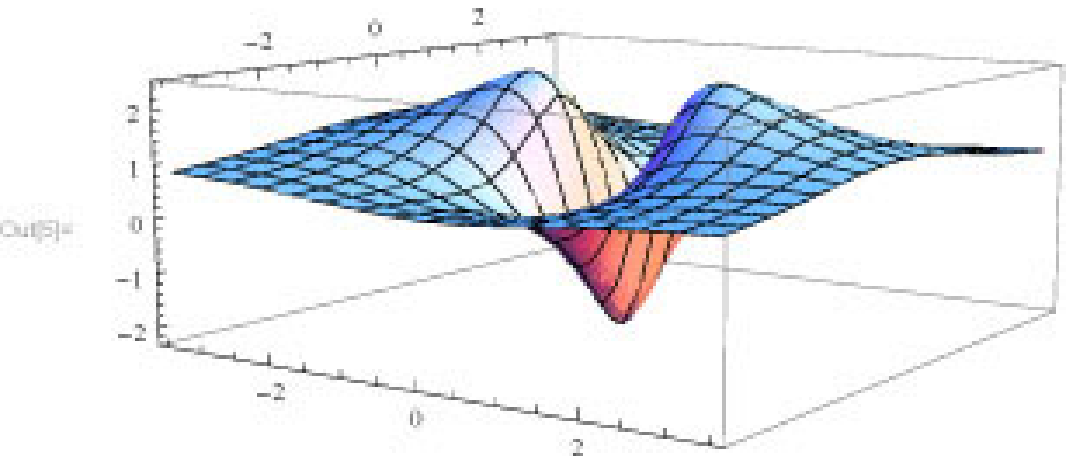
,height=8cm,width=10cm}
} \caption{$XYZ$ average energy in maximally entangled $|G_+\ra$ state for $ J_+=-1.5, J_-= -1.5, J_z=1.5$}
\end{figure}

For the state $| G_-\ra$ it is independent of $\psi$:
\be \la G_- | H| G_-\ra= -(\frac{J_z}{2}+ J_+). \ee

\subsection{Three Qubit case XYZ model}
Now we consider  three qubit coherent state
\be |PG_+\ra= \frac{1}{\sqrt{2}}(|\psi\ra |\psi\ra |\psi\ra + |-\psi^*\ra |-\psi^*\ra |-\psi^*\ra) \ee

This state can be obtained from maximally entangled GHZ state
\be |GHZ\ra= \frac{1}{\sqrt{2}} (|000\ra+|111\ra) \ee
by unitary transformation $U=U\otimes U\otimes U.$ This state is also maximally entangled and in the special case $\psi \rightarrow 0$ and $\psi^*\rightarrow \infty $ reduces to the GHZ state.

Then we have energy
\be
\la PG_+|H |PG_+ \ra = \frac{4 J_+ |\psi|^2 (1 + |\psi|^2) +  2 J_- (1+ |\psi|^2) (\psi^2 + \bar\psi^2) + J_z (1 -  |\psi|^2 -  |\psi|^4 + |\psi|^6)}{ (1 + |\psi|^2)^3}
 \ee
 It is shown in Fig.6 and has four local extrema points with two maxima and two minima.

\begin{figure}[htbp]
\centerline{\epsfig{figure=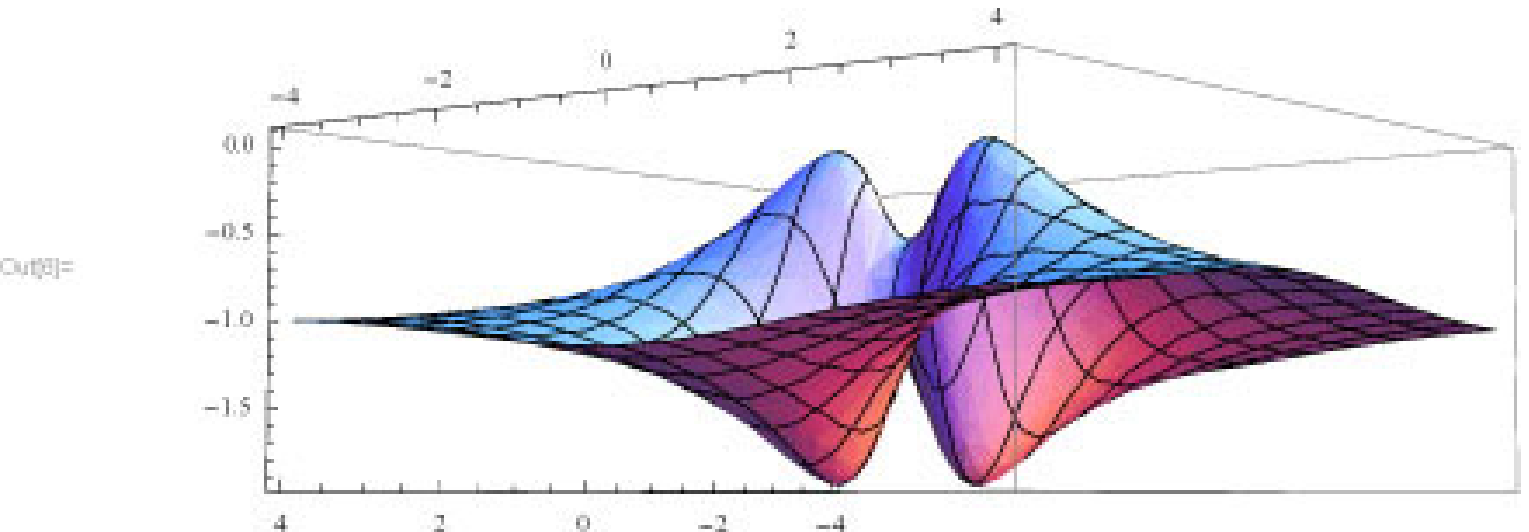
,height=8cm,width=10cm}
} \caption{$XYZ$ average energy in maximally entangled $|P G_+\ra$ state for $ J_+=-1, J_-= -1, J_z=-1$}
\end{figure}

\begin{figure}[htbp]
\centerline{\epsfig{figure=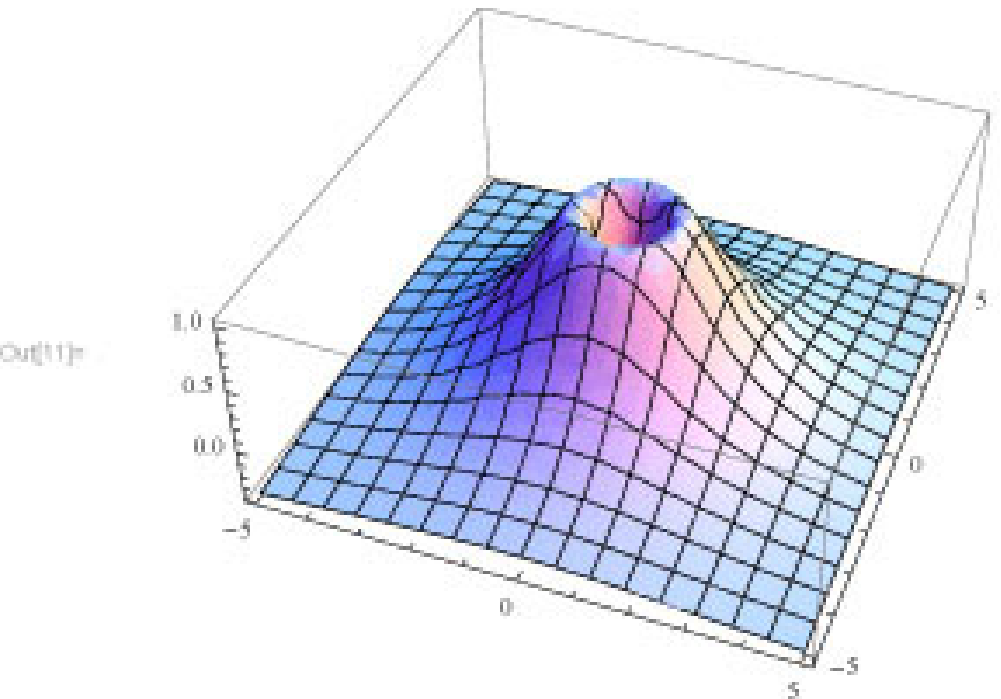
,height=8cm,width=10cm}
} \caption{$XYZ$ average energy in maximally entangled $|P G_+\ra$ state for $ J_+=1, J_-= 0, J_z=-0.5$}
\end{figure}
Another three qubit coherent state

\be |PG_- \ra = \frac{1}{\sqrt{3}}\left( |\psi \ra |\psi \ra |-\psi^*\ra  +   |\psi\ra |-\psi^*\ra |\psi\ra    +    |-\psi^*\ra |\psi\ra |\psi\ra  \right)
\ee

is related with maximally entangled $ |W \ra$ state

\be |W \ra = \frac{1}{\sqrt{3}}\left( |0 \ra |0 \ra |1\ra  +   |0\ra |1\ra |0\ra    +    |1\ra |0\ra |0\ra  \right)
\ee

For energy in this state we have

\be
\la PG_-|H |PG_- \ra = \frac{4 J_+ (1 + |\psi|^6) - 6 J_- (1+ |\psi|^2) (\psi^2 + \bar\psi^2) - J_z (1 - 9 |\psi|^2 - 9 |\psi|^4 + |\psi|^6)}{3 (1 + |\psi|^2)^3}
 \ee
\begin{figure}[htbp]
\centerline{\epsfig{figure=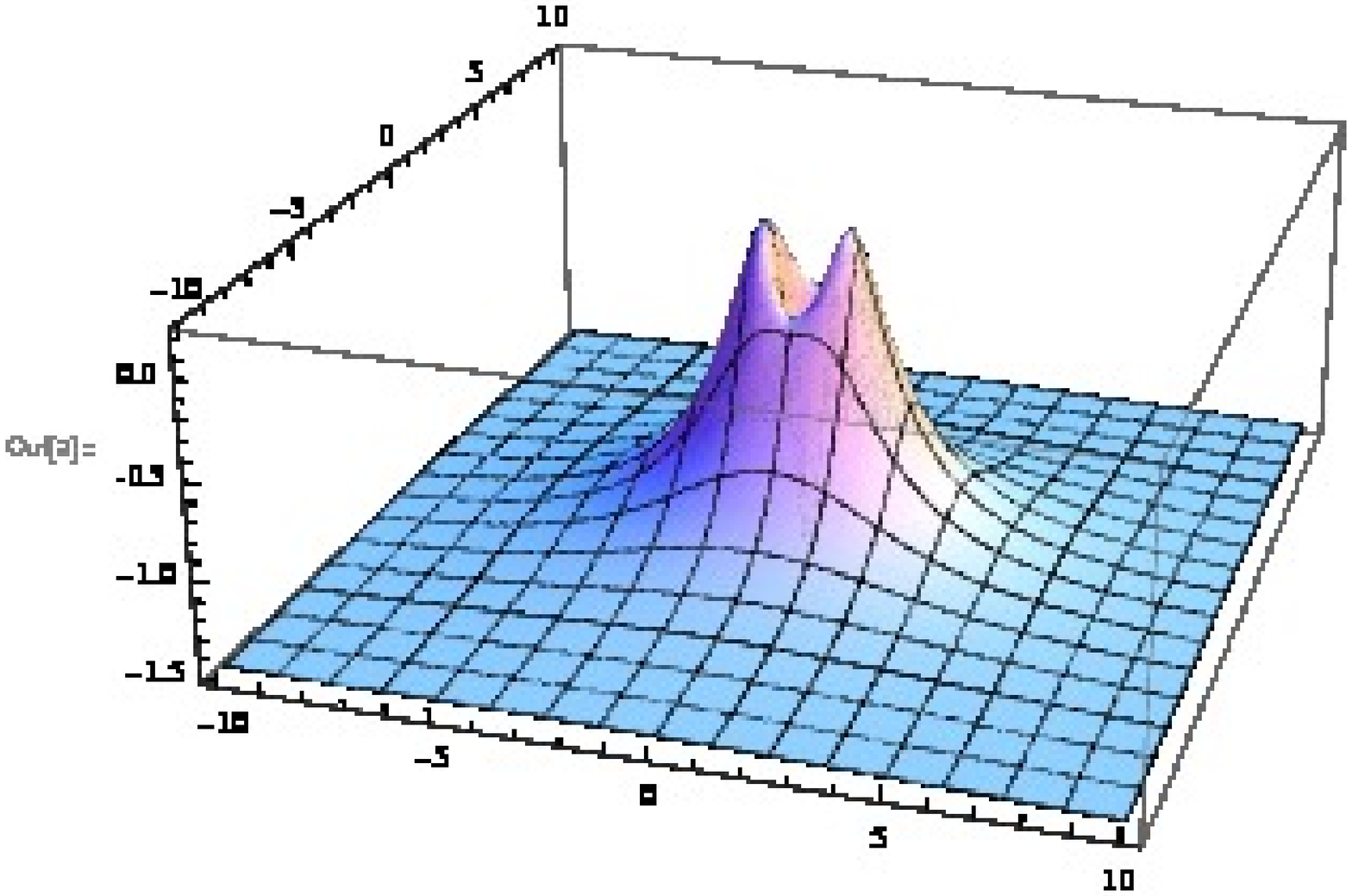
,height=8cm,width=10cm}
} \caption{$XYZ$ average energy in maximally entangled $|P G_-\ra$ state for $ J_+=-1, J_-= -0.2, J_z=0.5$}
\end{figure}

It is shown in Fig.7.

All above figures show the average energy as function of $\psi$ for maximally entangled coherent states and can be interpreted as soliton configurations in q-bit parametric space.
Then, depending on number of qubits and parameters of the system we can identify local extremum points as multisoliton configurations.

\section{Time Evolution}
Here we study time evolution of two qubit state prepared as the coherent state. The evolution
operator $$ U(t)= \exp [ -\frac{i}{\hbar} H t]$$ is determined by two
qubit Hamiltonian of $XYZ$ model.

%Using separable coherent state as initial state we find evolution in the form \be |\psi\ra = |\psi_1\,\, 0 \ra  \ee \be U(t)|\psi_1\,\, 0\ra = A_{11} |00\ra + %\frac{\psi_1}{\sqrt{1+|\psi_1|^2}} A_{23}|01\ra+ \frac{\psi_1}{\sqrt{1+|\psi_1|^2}}  A_{33}|01\ra + A_{41}|11\ra \ee
%The concurrence
%for this state $U(t)|\psi_1\,\, 0\ra$ is
%\bea C&=& 2 |t_0 t_3 - t_1 t_2|\\
%&=& 2 \left|A_{11}A_{41}- A_{23} A_{33} \frac{\psi_1^2}{1+|\psi_1|^2} \right|\eea
%For another initial state $|\psi \ra = (|\psi_1 \ra + |\psi_1^* \ra) |0 \ra  $
%we have the concurrence as
%\be C= 2 \left|A_{11}A_{41} \frac{(1+ |\psi_1|)^2}{1+|\psi_1|^2}- A_{23} A_{33} \frac{|\psi_1|^2}{(\bar{\psi}_1)^2 (1+|\psi_1|^2)} \right| \ee
Here for simplicity we display particular case of $XX$ model.
If we choose initial state as maximally entangled coherent state $|P_+\ra$, then we get evolution
\bea | \psi(t)\ra &=&  U|P_+\ra \\
&=&\frac{1}{\sqrt{2}(1+|\psi|^2)}[(1+\bar{\psi}^2)|00\ra+ e^{\frac{-i J t}{\hbar}}(\psi-\bar{\psi})|01\ra \nonumber \\&+& e^{\frac{-iJ t}{\hbar}}(\psi-\bar{\psi})|10\ra+ (1+\psi^2)|11\ra] \nonumber. \eea
For concurrence we have time dependence
 \bea C(t)&=& 2 |t_{00}t_{11}-t_{01}t_{10}|\\
&=& \left| \frac{(1+\bar{\psi}^2)(1+\psi^2)- e^{-\frac{2i Jt}{\hbar}}(\psi- \bar{\psi})^2}{(1+\bar{\psi}^2)^2} \right|
%&=& 1+ 2 \cos{(\frac{2 J t }{\hbar}- 2\theta)} \tan^2{\theta}+ tan^4{\theta}
. \eea
For  $ \psi= e^{i \theta}$ it gives
 \be C(t)= \frac{1}{4}\sqrt{(2+2\cos{2 \theta}^2)^2 + 8 (2+2\cos{2 \theta}^2) \sin^2{\theta}\cos{\frac{2 J t}{\hbar}}+ 16 \sin^4{\theta} } .\ee
 \begin{figure}[htbp]
\centerline{\epsfig{figure=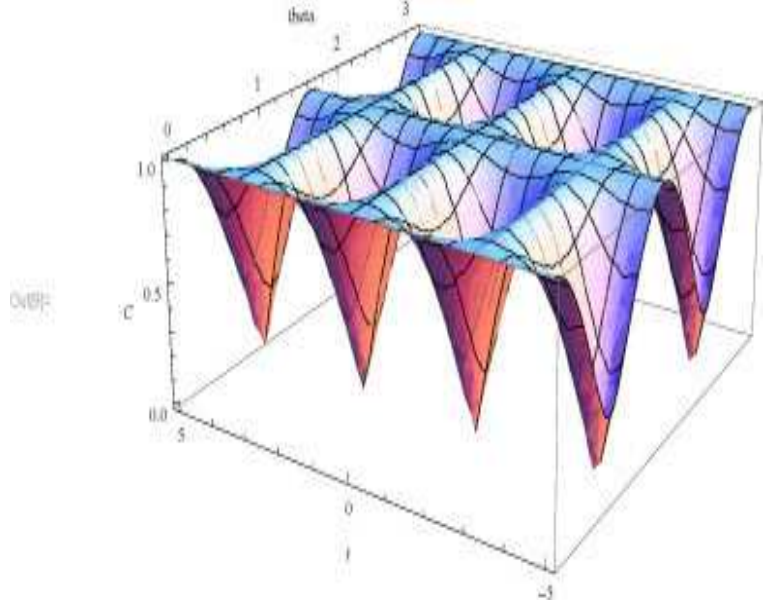
,height=8cm,width=10cm}
} \caption{Concurrence versus Time $J=1$}
\end{figure}

and for the fidelity \bea F(t)&=& |\la \psi(t)| P_+\ra|^2 \\
&=& \left|   \frac{|(1+\bar{\psi}^2|^2+ e^{\frac{i J t}{\hbar}}|\psi-\bar{\psi}|^2}{(1+|\psi|^2)^2} \right|^2.
\eea By parameterizations
$|\psi|^2=1 \rightarrow  \psi= e^{i \theta}$ we have
\be F(t)= 1- \sin^2{2 \theta} \sin^2{\frac{J t}{\hbar}}.\ee
\begin{figure}[htbp]
\centerline{\epsfig{figure=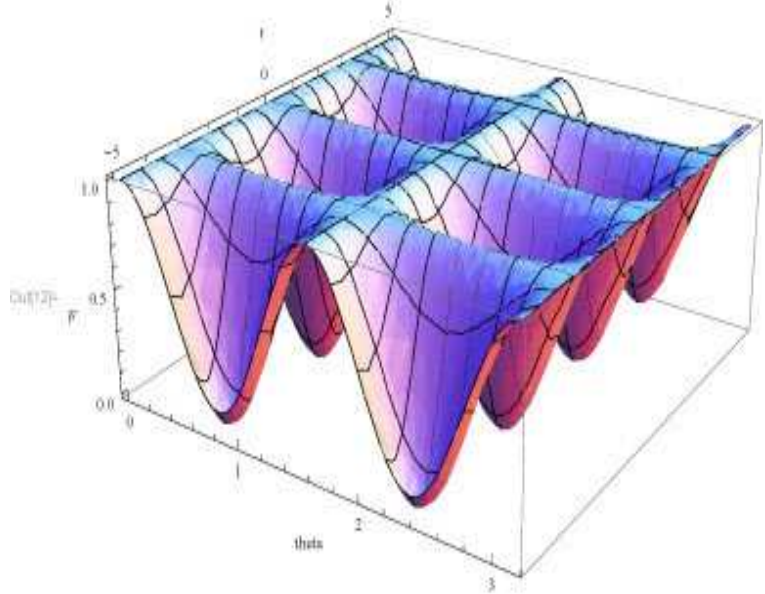
,height=8cm,width=10cm}
} \caption{Fidelity versus Time}
\end{figure}

This show that at time $t=\frac{2 \pi \hbar}{J}$, $n=0,1,2,...$ the evolved state return back to maximally entangled coherent state $|P_+\ra.$ We display this evolution in Fig.9 and Fig.10.

\section{CONCLUSIONS}
\hskip 0.5cm
In the present paper we introduced the set of maximally entangled two and three qubit coherent states determined by antipodal points on Bloch spheres. In complex plane these states are related with inverse symmetrical points under unit circle and can be interpreted as a some type of source and image similar to hydrodynamic vortex case \cite{Pashaevyilmaz}. From this point of view our coherent states realized method of images applied to quantum mechanical problem.
Some results on relation of M\"obius transformation to one qubit states were published in \cite{Lee}. Very recently special type of permutation symmetrical states and their representation on the Majorana sphere and action of M\"obius transformation on these states were discussed. But our approach and results are different from these papers.
Moreover, our procedure can be extended to construct multi qubit coherent states. Interesting question here is constructing the average energy as a function of phase space variables. We expect that the energy surface in multi qubit case will show specific multi soliton type of structure where number of solitons would be connected with the number of qubits. This question is under investigation.

\section{Acknowledgments}
 This work has been supported by the National Research Foundation, Ministry of Education, Singapore and Izmir Institute of Technology, Turkey.

\end{document}